\documentstyle[12pt]{article}

\begin{document}

\title{Generalizing Elitzur-Vaidman interaction free measurements}

\author{Adonai S. Sant'Anna \and Ot\'avio Bueno}

%\altaffiliation{Permanent address: Departamento de Matem\'atica,
%UFPR, C.P. 019081, Curitiba, PR, 81531-990, Brazil}

%\email{adonai@ufpr.br}

%\author{Ot\'avio Bueno}

%\email{obueno@sc.edu}

\date{Department of Philosophy, University of South Carolina,
Columbia, SC, 29208\\adonai@ufpr.br}

%\begin{document}

%\renewcommand{\thesection}{\Roman{section}}
\newtheorem{definicao}{Definition}
\newtheorem{teorema}{Theorem}
\newtheorem{lema}{Lemma}
\newtheorem{corolario}{Corolary}
\newtheorem{proposicao}{Proposition}
\newtheorem{axioma}{Axiom}
\newtheorem{observacao}{Observation}

%\maketitle

%\newpage

%\pacs{03.65.Ta, 03.65.Ud, 03.75.-b}

%\keywords{Foundations of quantum mechanics, interaction free
%measurement, matter waves}

\maketitle

\begin{abstract}

In the early 1990's A. Elitzur and L. Vaidman proposed an
interaction free measurement (IFM) that allows researchers to find
infinitely fragile objects without destroying them. But
Elitzur-Vaid\-man IFM has been used only to determine the position
of opaque objects. In this paper, we propose an extension of such
a technique that allows measurement of classical electric and
magnetic fields. Our main goal is to offer a framework for future
investigations about the role of the measurement processes,
expanding the physical properties that are measurable by means of
IFM.

\end{abstract}

\section{Introduction}

Avshalom Elitzur and Lev Vaidman proposed a technique for an
interaction free measurement (IFM) that allows researchers to find
infinitely fragile objects without destroying them
\cite{Elitzur-93}. Some experimental demonstrations of this
prediction have been obtained \cite{Kwiat-95,Kwiat-99}. For a
recent review on the theoretical and experimental aspects of the
IFM proposed by Elitzur and Vaidman, see \cite{Vaidman-03}.

Nevertheless, Elitzur-Vaidman IFM has been used only to determine
the position of non-transparent objects. The original scheme is
very simple. It is based on a Mach-Zehnder interferometer (see
FIG. 1). Single photons are emitted to the first beam splitter
(BS$_1$) with a transmission coefficient 1/2. Next, the
transmitted and reflected parts of the photon wave are reflected
by mirrors M$_1$ and M$_2$, respectively. These reflected waves
are reunited at the beam splitter BS$_2$, whose transmission
coefficient is 1/2 as well. Two photon detectors, LD (light
detector) and DD (dark detector), are positioned according to FIG.
1. The geometric arrangement is made in such a way that all
photons are detected at LD and no photon is detected at DD, due to
the self-interference of photons.

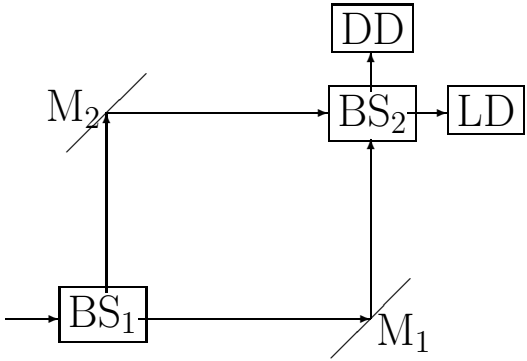
\begin{figure}

\begin{picture}(250,135.0)(1.0,1.0)

%\linethickness{1mm}

\put(5,35){\vector(1,0){20}}

\put(55,35){\vector(1,0){88}}

\put(43,45){\vector(0,1){68}}

\put(43,113){\vector(1,0){84}}

\put(157,113){\vector(1,0){15}}

\put(143,35){\vector(0,1){68}}

\put(143,121){\vector(0,1){15}}

\put(25,32){\framebox{\Large BS$_1$}}

\put(127,108){\framebox{\Large BS$_2$}}

\put(172,108){\framebox{\Large LD}}

\put(128,139){\framebox{\Large DD}}

\put(20,110){\Large M$_2$}

\put(145,25){\Large M$_1$}

\put(28,98){\line(1,1){30}}

\put(128,20){\line(1,1){30}}

\end{picture}

\vskip-5mm

\caption{Mach-Zehnder interferometer.}

\end{figure}

If any opaque object -- like, for example, an infinitely fragile
object that explodes when it is hit by any photon -- is put on the
way, say, between BS$_1$ and M$_2$, then there is no interference
phenomenon. In this case, there is a 25\% chance that DD detects a
single photon sent through the interferometer. If a single photon
is detected at DD after it was sent to the interferometer, then we
can know for sure that there is something inside the
interferometer. This is called an interaction free measurement in
the sense that there is a 25\% probability of knowing that there
is in fact a photon-sensitive bomb inside the interferometer
without exploding it. The fact that that bomb is found in a region
of space without exploding it is evidence that there was no
interaction between the photon detected at DD and the bomb.
Obviously, if a photon is detected at LD, we know nothing at all
about any object inside the interferometer. Besides, there is
still a 50\% chance of any single photon emitted to BS$_1$ be
reflected and hit the bomb. In this case, we find the bomb with an
interaction measurement, since it actually explodes.

In \cite{Vaidman-03} Vaidman makes a very clear review of the
meaning of the term ``IFM''. In \cite{Kwiat-95} the authors
demonstrate this technique in laboratory and still make an
improvement to increase the efficiency of the scheme, using a
discrete form of the quantum Zeno effect.

In this paper, we propose an extension of such a technique
allowing researchers to measure electric and magnetic fields
generated by macroscopic sources. In principle, even gravitational
fields are measurable by an analogous experiment. One of the main
differences between Elitzur-Vaidman proposal and ours is the use
of matter waves instead of photon waves.

We believe that Elitzur-Vaidman's approach as well as the {\em
Gedanken\/} experiments we propose in this paper are an important
part of a more comprehensive understanding of the nature of
measurement processes. But to this effect, it is important to
determine the physical properties that are measurable by IFM, and
those that are not.

\section{Matter waves}

There is no trivial extension of the optical elements of the
Mach-Zehnder interferometer for matter waves like neutrons,
electrons, atoms, and molecules. This is due to the fact that
there is no such thing as mirrors and beam splitters for massive
particles. In this case, the apparatus that has the best
similarity to a Mach-Zehnder interferometer is the three-grating
Mach-Zehnder interferometer, which is described in FIG. 2.

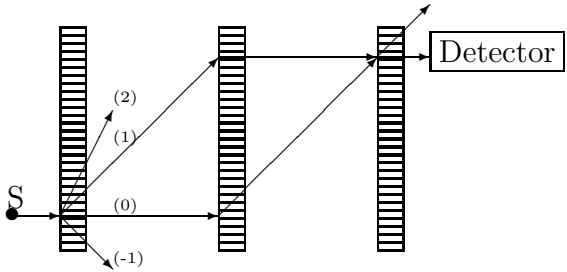
\begin{figure}

\begin{picture}(190,130.0)(1.0,1.0)

\put(0,20){\framebox{$\;$}} \put(0,23){\framebox{$\;$}}
\put(0,26){\framebox{$\;$}} \put(0,29){\framebox{$\;$}}
\put(0,32){\framebox{$\;$}} \put(0,35){\framebox{$\;$}}
\put(0,38){\framebox{$\;$}} \put(0,41){\framebox{$\;$}}
\put(0,44){\framebox{$\;$}} \put(0,47){\framebox{$\;$}}
\put(0,50){\framebox{$\;$}} \put(0,53){\framebox{$\;$}}
\put(0,56){\framebox{$\;$}} \put(0,59){\framebox{$\;$}}
\put(0,62){\framebox{$\;$}} \put(0,65){\framebox{$\;$}}
\put(0,68){\framebox{$\;$}} \put(0,71){\framebox{$\;$}}
\put(0,74){\framebox{$\;$}} \put(0,77){\framebox{$\;$}}
\put(0,80){\framebox{$\;$}} \put(0,83){\framebox{$\;$}}
\put(0,86){\framebox{$\;$}} \put(0,89){\framebox{$\;$}}
\put(0,92){\framebox{$\;$}} \put(0,95){\framebox{$\;$}}
\put(0,98){\framebox{$\;$}}

\put(60,20){\framebox{$\;$}} \put(60,23){\framebox{$\;$}}
\put(60,26){\framebox{$\;$}} \put(60,29){\framebox{$\;$}}
\put(60,32){\framebox{$\;$}} \put(60,35){\framebox{$\;$}}
\put(60,38){\framebox{$\;$}} \put(60,41){\framebox{$\;$}}
\put(60,44){\framebox{$\;$}} \put(60,47){\framebox{$\;$}}
\put(60,50){\framebox{$\;$}} \put(60,53){\framebox{$\;$}}
\put(60,56){\framebox{$\;$}} \put(60,59){\framebox{$\;$}}
\put(60,62){\framebox{$\;$}} \put(60,65){\framebox{$\;$}}
\put(60,68){\framebox{$\;$}} \put(60,71){\framebox{$\;$}}
\put(60,74){\framebox{$\;$}} \put(60,77){\framebox{$\;$}}
\put(60,80){\framebox{$\;$}} \put(60,83){\framebox{$\;$}}
\put(60,86){\framebox{$\;$}} \put(60,89){\framebox{$\;$}}
\put(60,92){\framebox{$\;$}} \put(60,95){\framebox{$\;$}}
\put(60,98){\framebox{$\;$}}

\put(120,20){\framebox{$\;$}} \put(120,23){\framebox{$\;$}}
\put(120,26){\framebox{$\;$}} \put(120,29){\framebox{$\;$}}
\put(120,32){\framebox{$\;$}} \put(120,35){\framebox{$\;$}}
\put(120,38){\framebox{$\;$}} \put(120,41){\framebox{$\;$}}
\put(120,44){\framebox{$\;$}} \put(120,47){\framebox{$\;$}}
\put(120,50){\framebox{$\;$}} \put(120,53){\framebox{$\;$}}
\put(120,56){\framebox{$\;$}} \put(120,59){\framebox{$\;$}}
\put(120,62){\framebox{$\;$}} \put(120,65){\framebox{$\;$}}
\put(120,68){\framebox{$\;$}} \put(120,71){\framebox{$\;$}}
\put(120,74){\framebox{$\;$}} \put(120,77){\framebox{$\;$}}
\put(120,80){\framebox{$\;$}} \put(120,83){\framebox{$\;$}}
\put(120,86){\framebox{$\;$}} \put(120,89){\framebox{$\;$}}
\put(120,92){\framebox{$\;$}} \put(120,95){\framebox{$\;$}}
\put(120,98){\framebox{$\;$}}

\put(-20,30){\vector(1,0){20}}

\put(0,30){\vector(1,1){60}}

\put(0,30){\vector(1,2){20}}

\put(60,30){\vector(1,1){60}}

\put(60,90){\vector(1,0){60}}

\put(0,30){\vector(1,-1){20}}

\put(120,90){\vector(1,1){20}}

\put(0,30){\vector(1,0){60}}

\put(120,90){\vector(1,0){20}}

\put(20,32){\tiny (0)}

\put(20,58){\tiny (1)}

\put(20,73){\tiny (2)}

\put(20,12){\tiny (-1)}

\put(140,88){\framebox{Detector}}

\put(-20,33){S}

\put(-21,27.5){$\bullet$}

\end{picture}

\caption{Three-grating Mach-Zehnder interferometer.}

\end{figure}

The optical elements of the Mach-Zehnder interferometer are all
replaced by either crystals or nanofabricated diffraction
membranes \cite{Keith-91} or even laser standing waves
\cite{Buchner-03}. The idea is to consider diffraction instead of
beam splitting and reflection.

A collimated matter wave is emitted from a source S, according to
FIG. 2. When this wave hits the first grating, the beam is
diffracted in several diverging orders, where the primary ones are
-1, 0, and 1. The 0$^{th}$ and the 1$^{st}$ order beams are
diffracted through the second grating. The second grating
diffracts a portion of each of the two incoming beams towards each
other. Such diffracted beams, which are the 1$^{st}$ and the
-1$^{st}$ orders of the two incident beams, respectively, overlap
at the third grating. A resulting beam is then measured by a
detector, which can be, for example, a wire that is much wider
than a grating period. In other words, the first and the third
grating play the role of the beam splitters in the traditional
Mach-Zehnder interferometer, while the second grating plays the
role of the two mirrors (FIG. 1).

One of the fundamental aspects of the IFM proposed by Elitzur and
Vaidman is the fact that photons are emitted one by one. Matter
waves interferometry in many cases is not performed with single
particles. One rare exception is the electron interferometry with
single electrons performed by Akira Tonomura \cite{Tonomura-03}.
Nevertheless, such an interferometry is accomplished in a two-slit
type experiment, which is not geometrically equivalent to a
Mach-Zehnder interferometer. In our {\em Gedanken\/} experiment
proposed in the next section, it is fundamental that we have a
certain distance between two primary paths of a diffracted beam,
much bigger than the usual in a two-slit experiment.

There has been a remarkable improvement in the technology of
matter waves interferometry. But some advances are still needed
for single particle interferometers. One of the difficulties for
this kind of technology is a lack of suitable diffraction elements
for manipulating coherent atomic and molecular de Broglie waves.
But we hope that our main ideas here will be testable in the
future, mainly by means of the improvement of waves matter
interferometry.

\section{{\em Gedanken\/} experiments}

Consider a three-grating Mach-Zehnder interferometer with a
macroscopic source C of classical fields in the vicinity of the
interferometer, which is illustrated in FIG. 3.

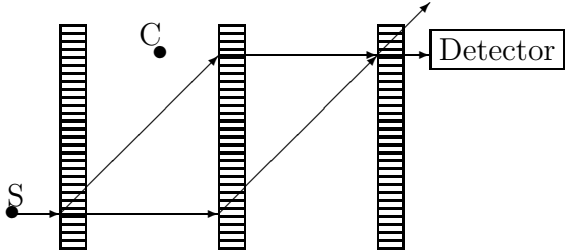
\begin{figure}

\begin{picture}(190,130.0)(1.0,1.0)

\put(0,20){\framebox{$\;$}} \put(0,23){\framebox{$\;$}}
\put(0,26){\framebox{$\;$}} \put(0,29){\framebox{$\;$}}
\put(0,32){\framebox{$\;$}} \put(0,35){\framebox{$\;$}}
\put(0,38){\framebox{$\;$}} \put(0,41){\framebox{$\;$}}
\put(0,44){\framebox{$\;$}} \put(0,47){\framebox{$\;$}}
\put(0,50){\framebox{$\;$}} \put(0,53){\framebox{$\;$}}
\put(0,56){\framebox{$\;$}} \put(0,59){\framebox{$\;$}}
\put(0,62){\framebox{$\;$}} \put(0,65){\framebox{$\;$}}
\put(0,68){\framebox{$\;$}} \put(0,71){\framebox{$\;$}}
\put(0,74){\framebox{$\;$}} \put(0,77){\framebox{$\;$}}
\put(0,80){\framebox{$\;$}} \put(0,83){\framebox{$\;$}}
\put(0,86){\framebox{$\;$}} \put(0,89){\framebox{$\;$}}
\put(0,92){\framebox{$\;$}} \put(0,95){\framebox{$\;$}}
\put(0,98){\framebox{$\;$}}

\put(60,20){\framebox{$\;$}} \put(60,23){\framebox{$\;$}}
\put(60,26){\framebox{$\;$}} \put(60,29){\framebox{$\;$}}
\put(60,32){\framebox{$\;$}} \put(60,35){\framebox{$\;$}}
\put(60,38){\framebox{$\;$}} \put(60,41){\framebox{$\;$}}
\put(60,44){\framebox{$\;$}} \put(60,47){\framebox{$\;$}}
\put(60,50){\framebox{$\;$}} \put(60,53){\framebox{$\;$}}
\put(60,56){\framebox{$\;$}} \put(60,59){\framebox{$\;$}}
\put(60,62){\framebox{$\;$}} \put(60,65){\framebox{$\;$}}
\put(60,68){\framebox{$\;$}} \put(60,71){\framebox{$\;$}}
\put(60,74){\framebox{$\;$}} \put(60,77){\framebox{$\;$}}
\put(60,80){\framebox{$\;$}} \put(60,83){\framebox{$\;$}}
\put(60,86){\framebox{$\;$}} \put(60,89){\framebox{$\;$}}
\put(60,92){\framebox{$\;$}} \put(60,95){\framebox{$\;$}}
\put(60,98){\framebox{$\;$}}

\put(120,20){\framebox{$\;$}} \put(120,23){\framebox{$\;$}}
\put(120,26){\framebox{$\;$}} \put(120,29){\framebox{$\;$}}
\put(120,32){\framebox{$\;$}} \put(120,35){\framebox{$\;$}}
\put(120,38){\framebox{$\;$}} \put(120,41){\framebox{$\;$}}
\put(120,44){\framebox{$\;$}} \put(120,47){\framebox{$\;$}}
\put(120,50){\framebox{$\;$}} \put(120,53){\framebox{$\;$}}
\put(120,56){\framebox{$\;$}} \put(120,59){\framebox{$\;$}}
\put(120,62){\framebox{$\;$}} \put(120,65){\framebox{$\;$}}
\put(120,68){\framebox{$\;$}} \put(120,71){\framebox{$\;$}}
\put(120,74){\framebox{$\;$}} \put(120,77){\framebox{$\;$}}
\put(120,80){\framebox{$\;$}} \put(120,83){\framebox{$\;$}}
\put(120,86){\framebox{$\;$}} \put(120,89){\framebox{$\;$}}
\put(120,92){\framebox{$\;$}} \put(120,95){\framebox{$\;$}}
\put(120,98){\framebox{$\;$}}

\put(-20,30){\vector(1,0){20}}

\put(0,30){\vector(1,1){60}}

\put(60,30){\vector(1,1){60}}

\put(60,90){\vector(1,0){60}}

\put(120,90){\vector(1,1){20}}

\put(0,30){\vector(1,0){60}}

\put(120,90){\vector(1,0){20}}

\put(140,88){\framebox{Detector}}

\put(-20,33){S}

\put(-21,27.5){$\bullet$}

\put(35,88){$\bullet$}

\put(30,94){C}

\end{picture}

\caption{Our proposed {\em Gedanken\/} experiment with the source
C of classical fields.}

\end{figure}

Suppose that the other source S is capable of emitting coherent
single particles, and that the detector that is put after the
third grating is capable of detecting these single particles.

We can put the third grating in a position where we expect no
particles at all in the detector, due to the destructive effect of
the interference fringes. Of course, this is an ideal situation
for now, since there is no technology at the present that
accomplishes this with point electric charges. We call this the
``ideal condition''.

Our idea is to measure the field that is emitted by a source C but
without any interaction with the field. Our proposal works for a
field that has enough intensity to disturb any particle of the
diffracted beam of order 1 but weak enough to have its physical
effects on the beam of order 0 completely neglected. If the source
C is, e.g., a point electric charge, then the matter wave can be
formed by either single electrons or single ions.

As is well known \cite{Matveyev-66}, any point charge submitted to
an electromagnetic field will experience a quantum mechanical
version of the Lorentz force, given by ${\bf F} = q [ {\bf
E}+\frac{1}{2c}\left({\bf v}\times {\bf B} - {\bf B}\times {\bf
v}\right) ]$, where $c$ is the speed of light in the vacuum, $q$
is the electric charge of the point charge, ${\bf v}$ is its speed
and ${\bf E}$ and ${\bf B}$ are the electric and the magnetic
components, respectively, of the electromagnetic field. We assume
here that the point charge corresponds to the single particles
emitted by S, while ${\bf E}$ and ${\bf B}$ may be generated by C.

Now we consider the particular case of an electric field, constant
in time and emitted by C, strong enough to affect any particle in
a given vicinity, but weak enough to be neglected outside this
vicinity. Besides, this field exerts an attractive force over
particles emitted by S.

Let us suppose that we can control the distance of C to the
expected path of the diffracted beam of order 1, in a way that C
is always at a bigger distance from the expected path of the
diffracted beam of order 0. Moreover, C has no movement during the
IFM process.

Let us also suppose that the size of the second grating is quite
limited in the following way: some attractive forces (into the
direction of C) are supposed to bend the expected trajectory of a
given particle at a certain speed, causing a deflection of an
angle above a critical value $\varphi_c$. Any deflection angle
above $\varphi_c$ corresponds to a particle that never reaches the
second grating, but it follows away from the interferometer, due
to the appropriate limitation on the size of that grating.

So, if C is distant enough, then the effect of its generated
electric field over the interferometer can be neglected and no
particles are expected to be registered at the detector, due to
the ``ideal condition'' mentioned above. But when we discretely
move C (a movement during a time interval where no charge particle
emitted by S is inside the interferometer) toward the expected
path of the first beam of order 1, it will happen at a certain
point that the field is strong enough to bend the trajectory of
any charged particle of this beam. We propose a discrete movement
of C to avoid accelerations that could generate a magnetic
component associated to the electric field, during the time
interval in which the particle is inside the interferometer. If
this distance between C and the interferometer reaches a value
that goes beyond a critical value that is responsible for a
deflection angle greater than $\varphi_c$, then there is a
non-null probability of registering single particles at the
detector. This means that if any particle is registered at the
detector, then we know that the electric field generated by C is
above a critical value, which corresponds to a measurement of such
an electric field within an error that depends on the discrete
steps used to move C. It is clear that for each discrete position
of C we must test the effects on the detector several times, until
we are sure that no particle will be registered at the detector.
We must repeat these tests to be sure that if no particle is
detected, then this is due to the fact that the field generated by
C is too weak to be measured.

On the other hand, if a particle is registered at the detector,
then we know that there is an electric field whose value at a
given point in the expected path of the diffracted beam of order 1
is greater than the critical value. Obviously, the interferometer
needs to be calibrated for each measurement. Since there is a
field source in C, there is a potential difference between any two
expected primary paths of the diffracted beam of charged particles
emitted by S. We know that such a potential difference may be
responsible for phase changes in the diffracted beams
\cite{Sakurai-94}, which are measurable in the detector. So, if a
particle is registered at the detector, then we need to be sure
that such a detection is not due to a constructive interference
phenomenon which may be explained by means of phase changes caused
by any potential difference associated to the electric field that
we intend to measure. So, we need to calibrate our measurement
device to isolate any effect due to such a potential difference.
After all, we are mainly interested on the electric field that is
responsible for a Lorentz force.

We calibrate the interferometer as follows. For a given position
of C, we put two metallic cages around the respective expected
paths of the two main beams that travel from the first to the
third grating. Any electric charge that travels through any one of
these cages will be subject to a constant potential inside the
cages, since inside the cages the potential is spatially uniform.
So, there will be no Lorentz force on these point charges,
although there is a potential difference between the expected
paths that may affect the interference pattern of the point
charges emitted by S. Next we adjust the position of the third
grating until no particle emitted by S is registered at the
detector. Now we are able to remove the cages in order to perform
the interaction free measurement of the field emitted by C. Such a
measurement is IFM in the sense described in the next paragraph.

If the trajectory of any electric charge may be bent causing a
deflection above a critical angle $\varphi_c$, then this bending
is caused by a Lorentz force with a null magnetic field. On the
other hand, the point particle that is supposed to have its
trajectory bent, generates an electric field by itself, which
exerts a Lorentz force over C. We take it that if a particle is
registered at the detector of the interferometer, then no force is
necessarily registered over C. If C is a very fragile object that
cannot be disturbed by any force, even caused by a test particle,
then this is a good way to measure the field generated by C. This
also is an indirect measurement of the Lorentz force without any
counter-force involved. If any charged particle has its trajectory
bent by the electric field generated by C (the case with
interaction), then no particle is registered at the detector. But
if a particle is registered at the detector, we know for sure that
it had no interaction with the field. Moreover, the electric
particle registered in the detector has no change in its kinetic
energy, although it registers the approximate value of an electric
field (this value is above a critical value). This is the case
even though we know that the Lorentz force due to electric fields
always changes the velocity of electric charges under the
influence of this field. The point is that we make a measurement
of the field without any interaction with it. The measurement's
value is determined by the distance from C to the expected path of
the beam of order 1, which is responsible for the detection of the
electric charge.

The efficiency of our interaction free measurement is below the
25\% efficiency of the Elitzur-Vaidman proposal for IFM of
position. Let us assume that $p_1$ is the probability that any
charged particle follows from the first grating to the second one
without any diffraction (diffraction of 0$^{th}$ order). If $p_2$
is the probability that any charged particle follows from the
second grating to the third one with a 1$^{st}$ order diffraction,
the efficiency of our IFM is $p_1p_2$. Such efficiency depends on
the physical features of the gratings. One possible way to
increase this efficiency is by adapting the quantum Zeno effect
introduced in \cite{Kwiat-95} and \cite{Kwiat-99}.

An analogous framework could be used to measure magnetic fields.
We could replace C by a source of magnetic fields, without any
electric component. Despite the fact that magnetic fields do not
change the kinetic energy of any electric charge in movement, they
still exert a Lorentz force over such a point electric charge,
causing a bending in its expected path. If the trajectory of any
electric charge may be bent causing a deflection above a critical
angle $\varphi_c$, then this bending is caused by a Lorentz force
with a null electric field. On the other hand, the point particle
that is supposed to have its trajectory bent generates an electric
field by itself, which exerts a Lorentz force over C. In a similar
way, as described above, if a particle is registered at the
detector of the interferometer, then no force is registered over
C. In this sense, this measurement is interaction free if the
interferometer is obviously calibrated. The calibration is made in
a way similar to the case of the electric field measurement. The
goal of the calibration is to isolate our apparatus from any
physical effect of phase change in the emitted coherent particles
due to the vector potential associated to the magnetic field
(Aharonov-Bohm effect) \cite{Sakurai-94}.

An analogous experiment could be performed to measure, at least in
principle, the intensity of a gravitational field. This could be
done with interferometry of either heavy atoms or molecules, where
the mass of these particles is more relevant for gravitational
effects. Neutrons interferometry could be performed as well.
Nevertheless, we recognize that any conclusion about a true
interaction free nature of a measurement of this kind is not an
easy task.

Furthermore, technical difficulties are obviously greater if the
same strategy that uses photon interferometry is proposed to
measure gravitational fields. As is well known gravitational
fields are able to bend photon trajectories. The equation for
describing the first order term of the deflection of light under
the effect of a gravitational field caused by a mass $M$ is given
by $\Delta\phi = 4GM/(bc^2)$, where $G$ is a constant equal to
$6.67\times 10^{-8}cm^3g^{-1}s^{-2}$, $b$ is an impact parameter
and $c$ is a constant equal to $3.00\times 10^{10}cm\, s^{-1}$
\cite{Wald-84}. Suppose we want to detect a deflection angle of
$10^{-9}$ (a deflection of one nanometer to every one meter of
trajectory) in a photon trajectory that grazes the surface of a
massive sphere of, e.g., Iridium (one of the densest elements). In
this case, the sphere would need to have a radius of approximately
18,900 km. This is obviously an unrealistic situation, although it
is conceptually sound.

\section{Conclusion}

We gave the general framework of an IFM of some classical fields,
with special emphasis on the case of electric fields. In this
sense, we are extending the original Elitzur-Vaidman IFM, which
was designed to measure the position of an opaque object without
any interaction with it. Furthermore, we also show the advantages
that IFM with matter waves has with respect to the original scheme
proposed by Elitzur and Vaidman as well as some of the current
limitations of IFM in the measurement of gravitational fields.

\section{Acknowledgements}

We are grateful to Osvaldo Pessoa, Jr. and Jose C. Martinez for
helpful electronic discussions. This work was partially supported
by CAPES (Brazilian government agency) and by the National Science
Foundation (NSF 01-157, NIRT grant).

\end{document}